\documentstyle[multicol,aps,pre,epsf]{revtex}
\begin{document} 
\title{Persistence in the One-Dimensional $A+B \rightarrow \emptyset$
 Reaction-Diffusion Model} 
\author{S. J. O'Donoghue and A. J. Bray}
\address{Department of Physics and Astronomy, The University of Manchester,
M13 9PL, United Kingdom}

\date{\today}

\maketitle

\begin{abstract}

The persistence properties of a set of random walkers obeying the $A+B
\rightarrow  \emptyset$  reaction,   with  equal  initial  density  of
particles  and homogeneous  initial conditions,  is studied  using two
definitions  of   persistence.   The  probability,   $P(t)$,  that  an
annihilation  process  has  not  occurred  at  a  given  site  has  the
asymptotic form $P(t) \sim  {\rm const} + t^{-\theta}$, where $\theta$
is the persistence exponent  (``type I persistence'').  We argue that,
for a density of particles $\rho \gg 1$, this non-trivial
exponent is identical to  that governing the persistence properties of
the   one-dimensional  diffusion   equation,   $\partial_{t}  \phi   =
\partial_{xx}\phi$,  where $\theta \simeq  0.1207$ \cite{non}.  In the
case of  an initially low density,  $\rho_{0} \ll 1$,  we find $\theta
\simeq  1/4$  asymptotically.  The  probability that  a  site  remains
unvisited  by any  random  walker (``type  II  persistence'') is  also
investigated  and found to  decay with  a stretched  exponential form,
$P(t) \sim  \exp(- {\rm const} \times  \rho_0^{1/2}t^{1/4})$, provided 
$\rho_0 \ll 1$.  A heuristic argument for this behavior, based  on  an 
exactly solvable toy model, is presented. 
  
\end{abstract}

\begin{multicols}{2}

\section{Introduction}

Diffusion-limited  reactions  are  used  to  model  a  wide  range  of
phenomena in physics, chemistry and biology, and continue to stimulate
current  research. Reaction-diffusion processes  have been  applied to
studies such as  stochastic spin-flip dynamics \cite{0}, exciton-exciton
dynamics  in tetra-methylammonium  manganese trichloride  \cite{1}, the
kinetics  of  bipolymerisation  \cite{2},  reptation of  DNA  in  gels
\cite{3},  interface growth \cite{4},  diffusion of  zeolites \cite{5}
and   other  phenomena   such   as  competing   species  in   biology,
self-organised  criticality,  pattern   formation  and  dynamic  phase
transitions.

Much of the  effort to date has focused on reactions  of the form $A+A
\rightarrow \emptyset$ and $A+B \rightarrow \emptyset$, with a variety
of boundary  and initial conditions  \cite{0}. It is  well appreciated
that within the context of  these models there exists an upper critical
dimension  $d_{c}$, below  which spatial  fluctuations in  the initial
distribution of the reactants play a significant role in the evolution
of the  density of the  particles. This dependence on  the microscopic
fluctuations invalidates traditional approaches such as the mean-field
approximation. Attempts to understand  the role played by fluctuations
have   involved  numerous   techniques,   including  Smoluchowski-type
approximations \cite{6}  and field-theoretic methods  \cite{7,8,9}. In
this paper we  set out to elucidate the  effects of these fluctuations
on  \textit{persistence} within  the context  of the  $A+B \rightarrow
\emptyset$ model.
 
Persistence phenomena  have received considerable  attention in recent
years  \cite{non,10,11,12,13,14,15,16,17,18,19,20,21}. Theoretical and
computational  studies include  spin systems  in one  \cite{10,12} and
higher   \cite{13}   dimensions,   diffusion   fields   \cite{non,14},
fluctuating   interfaces  \cite{15}   and   phase  ordering   dynamics
\cite{16}.   Experimental studies include  the coarsening  dynamics of
breath figures  \cite{17}, soap  froths \cite{18} and  twisted nematic
liquid  crystals \cite{19}.   Persistence  in nonequilibrium  critical
phenomena has  also been  studied in the  context of the  global order
parameter     $M(t)$    regarded     as    a     stochastic    process
\cite{20}. Reaction-diffusion models offer much scope for the study of
persistence  and   have  already  contributed   significantly  to  our
understanding in this area \cite{21}.

The definition  of persistence  is as follows.   Let $\phi(x,t)$  be a
stochastic variable  fluctuating in space  and time according  to some
dynamics. The persistence probability is simply the probability $P(t)$
that at a  fixed point in space, the  quantity ${\rm sgn}\,[\phi(x,t)-
\langle\phi(x,t)\rangle]$  does not change  up to  time $t$.   In many
systems   of  physical   interest   a  power   law  decay,   $P(t)\sim
t^{-\theta}$, is observed, where  $\theta$ is the persistence exponent
and is, in general,  nontrivial. The nontriviality of $\theta$ emerges
as  a consequence  of the  coupling of  the field  $\phi(x,t)$  to its
neighbours, since such coupling implies that the stochastic process at
a fixed point in space and time is non-Markovian.

Two distinct types of persistence emerge naturally in the study of the
$A+B  \rightarrow  \emptyset$  reaction-diffusion model.   Consider  a
nonequilibrium field  $\phi(x,t)$ which  takes values at  each lattice
site $x$.  The field evolves in time $t$ through interactions with its
neighbours.  In type I  persistence, the field $\phi(x,t)$ changes its
sign whenever  an event occurs  at the lattice  site $x$ at  time $t$,
where an event is defined  to be the reaction process $A+B \rightarrow
\emptyset$. Type II  persistence satisfies the conventional definition
and the  field $\phi(x,t)$ changes sign  when the lattice  site $x$ is
visited by either an $A$ or $B$ particle at time $t$.  The persistence
probability at time  $t$ is defined as the fraction  of sites in which
the stochastic field $\phi(x,t)$ did  not change its value in the time
interval  $[0,t]$.  Our  analysis  suggests,  in the  case  of type  I
persistence,  that  when the  initial  density  of  particles is  high
($\rho_{0} \gg  1$, where $\rho_{0}$ is  the density of  either $A$ or
$B$ particles), the value of  the persistence exponent is identical to
that  which emerges  in  the study  of  the one-dimensional  diffusion
equation  $\partial_{t} \phi  = \partial_{xx}  \phi$, as  long  as the
running density,  $\rho(t)$, satisfies  $\rho(t) \gg 1$.   This yields
$\theta \simeq 0.1207$ \cite{non}.  When, however, the initial density
of particles is  low ($\rho_{0} \ll 1$), we  find $\theta \simeq 1/4$,
in agreement  with a simple heuristic  argument based on  the decay of
the density.   In the  case of  type II persistence  we show  that the
persistence decays with the  ``stretched exponential'' form $P(t) \sim
\exp(-  {\rm const}  \times  \rho_0^{1/2}t^{1/4})$, and  we provide  a
heuristic derivation of this based on the behavior of a toy model.

The  paper is organised  as follows.   In section  I we  introduce the
model  and calculate  the evolution  of the  particle density  for all
$\rho_{0}$, including $\rho_{0} \gg 1$.   In section II we present our
results  for  type I  persistence.   In  section  III we  introduce  a
generalised  toy  model  of  noninteracting  diffusing  particles  and
illuminate, through a special case, the type II persistence properties
of the  $A+B \rightarrow \emptyset  $ model.  All our  predictions are
tested by extensive numerical simulations.

\section{PARTICLE DENSITY IN THE $A+B \rightarrow \emptyset$ model}

We  are  concerned  with  the  following  model.   Consider  the  $A+B
\rightarrow \emptyset$ reaction involving  two types of particle, both
executing   diffusive  random   walks.   The   particles  move   on  a
one-dimensional lattice  with periodic boundary  conditions, and react
upon  contact to  form an  inert  particle.  At  $t=0$, exactly  equal
numbers, $N_{A}(0) = N_{B}(0)$, of $A$ particles and $B$ particles are
randomly  distributed  on  the  lattice.   This is  done  by  randomly
assigning one half of the lattice sites to $A$ particles and the other
half to $B$-particles.  The  $A$-sites and $B$-sites are then randomly
filled with $A$ and $B$ particles, i.e.\ each $A(B)$-particle occupies
each of the $A(B)$-sites with equal probability
such  that,   at  large  scales,  both   densities  $\rho_{A}(0)$  and
$\rho_{B}(0)$  are  initially  homogeneous.   We  define  $\rho_{0}  =
\rho_{A}(0) = \rho_{B}(0)  = N_{A}(0)/L$ where $L$ is  the size of the
lattice. The  two species are  also given the same  diffusion constant
$D_{A} = D_{B} = D =  1/2$. Our model then evolves permitting multiple
occupancy of  sites, but  we impose an  instantaneous reaction  so that
each lattice site contains only one type of particle.

The mean-field  approach yields an  asymptotic decay of  the particle
density  according  to $\rho_{A}(t)  =  \rho_{B}(t)  \sim {\rm  const}
\times  t^{-1}$   with  an   amplitude  independent  of   the  initial
density. However, in low enough  spatial dimension, $d \le d_{c} = 4$,
the dominating  process asymptotically is  the diffusive decay  of the
fluctuations  in  the  initial  conditions. This  leads  to  anomalous
kinetics and it was shown by Toussaint and Wilczek \cite{22} that
\begin{equation}
\left<   \rho_{A}(t)  \right>  =   \left<  \rho_{B}(t)   \right>  \sim
\frac{\sqrt\rho_{0}}{\pi^{1/2}(8\pi)^{d/4}}(Dt)^{-d/4} \ .
\label{tw}
\end{equation}
This result has been confirmed  using field-theoretic methods for $2 <
d <  4$ by Lee and Cardy  \cite{8}. The exponent $-d/4$  has also been
rigorously confirmed by Bramson and Lebowitz \cite{23}. The numerical
simulations  which  have  been  performed  in  one  \cite{22,24},  two
\cite{22,25} and three dimensions \cite{26} are also in good agreement
with the analytical predictions of the decay exponent. However, it was
noted  by Lee  and  Cardy  \cite{8} that,  although  in one  dimension
reasonable agreement  with the dependence  on the initial  density has
been  found,  in  higher  dimensions  the  $\sqrt{\rho(0)}$  amplitude
dependence  has  not  been   observed.   They  suggest  that,  in  the
one-dimensional simulations, the initial average occupation number per
site was kept  low, whereas for the higher  dimensional simulations it
was necessary  to start with a  nearly full lattice in  order to reach
the asymptotic regime and therefore that Eq. (\ref{tw}) might not be a
universal result,  but rather  a limit for  small initial  density. We
believe this to  be the case. We focus on the  particle density in one
dimension.  In  our study  of type I  persistence we  are particularly
interested in the limit of  high initial particle density, $\rho_0 \gg
1$,  where  the  problem  should  map onto  simple  diffusion.   As  a
precursor to looking at  the persistence, therefore, we first consider
how the particle density decays with time in this limit.

In the  high-density limit, where  there are many particles  per site,
one can neglect dynamical fluctuations  in the density. If $N_n(t)$ is
the  number of  particles at  site $n$  at time  $t$, the  dynamics of
$N_n(t)$ is governed by diffusion:
\begin{equation}
\dot{N}_{n}(t) = D(N_{n+1}(t) - 2N_{n}(t) + N_{n-1}(t))\ .
\label{diff}
\end{equation}
Here we have adopted the convention that $N_n > 0$ means that site $n$
is occupied  by $N_n$  $A$-particles, while $N_n<0$  means that  it is
occupied  by $-N_n$  $B$-particles. Then  the annihilation  process is
automatically   built  into   the  diffusion   equation  (\ref{diff}).
Introducing a discrete  Fourier transform and taking the  limit $L \to
\infty$, gives,
\begin{eqnarray}
N_{n}(t) & =  & \sum_{m}N_{m}(0) \int_{-\pi}^{\pi} \frac{dk}{2\pi}\exp
[-2D(1-\cos k)t \nonumber\\ && \hspace*{3cm} + ik(n-m)].
\end{eqnarray} 
For $t \gg 1$ we are justified  in using the approximation $1 - \cos k
\simeq  k^{2}/2$ in  the integrand,  and extending  the limits  on the
$k$-integral to $\pm \infty$. This gives
\begin{equation}
N_{n}(t)    =     \sum_{m}    N_{m}(0)    \frac{1}     {\sqrt{4    \pi
Dt}}\exp\left[-\frac{(n-m)^{2}}{4Dt}\right]\ .
\label{P(t)}
\end{equation}
For $t  \to \infty$, the  Gaussian kernel becomes broad,  and $N_n(t)$
becomes,  asymptotically,  a   Gaussian  random  variable  (where  the
randomness comes  from the initial  conditions). It's mean  is clearly
zero,  since   $\langle  N_m(0)  \rangle  =0$  by   symmetry,  so  its
distribution  is  completely  specified  by  its  variance.   This  is
independent of  $n$ by translational  invariance, so we  drop the
subscript and write (again for $t \to \infty$).
\begin{eqnarray}
\left<N^{2}(t)\right>         &         =        &         \sum_{m,m'}
\frac{\left<N_{m}(0)N_{m'}(0)\right>}{4\pi  Dt} \nonumber\\  && \times
\exp \left\{ -\frac{(n-m)^{2}}{4Dt}- \frac{(n-m')^{2}}{4Dt} \right\}.
\end{eqnarray}
Using   $\left<N_{m}(0)N_{m'}{0}\right>   =  \left<N_{m}^{2}(0)\right>
\delta_{mm'}$ we obtain, asymptotically,
\begin{equation}
\left<N^{2}(t)\right> = \frac{\left<N^{2}(0)\right>}{\sqrt{8\pi Dt}}.
\end{equation}

At time $t=0$ the number of particles  of a given type on a given site
has  a Poisson  distribution with  mean $2\rho_0$,  where  recall that
$\rho_{0}$ is half  the total density. Hence the  variance is given by
$\left<N^{2}(0)\right>  =  \left<|N(0)|^{2}\right>  = (2\rho_{0})^2  +
2\rho_{0}$, giving
\begin{equation}
\left<N^{2}(t)\right>  =  \frac{4\rho_{0}^{2} +  2\rho_{0}}{\sqrt{8\pi
Dt} } = \sigma^{2}(t)\ ,
\label{sigma}
\end{equation}
where  $\sigma$  is  the   standard  deviation,  and  the  probability
distribution of  $N(t)$ is simply $P[N(t)]  = (1/\sqrt{2\pi \sigma^2})
\exp(-N^{2}/2\sigma^{2})$.

The  total  particle density,  recalling  that,  with our  convention,
$B$-particles have been assigned negative values of $N_n(t)$, is given
by
\begin{equation}
\left<|N(t)|\right> = \sqrt{\frac{2}{\pi \sigma^{2}}}\int_{0}^{\infty}
dN N  \exp \left(-\frac{N^{2}}{2 \sigma^{2}}\right)  = \sqrt {\frac{2}
{\pi}}\sigma.
\end{equation} 
Substituting for $\sigma$ from Eq.(\ref{sigma}) yields, finally,
\begin{equation}
\left<|N(t)|\right> = \frac {(4  \rho_{0}^{2} + 2 \rho_{0})^{1/2}} {(2
\pi^{3} Dt)^{1/4}}.
\end{equation}
The  individual  particle  densities, therefore,  decay  asymptotically
according to,
\begin{equation}
\left<  \rho_{A}(t)  \right>   =  \left<  \rho_{B}(t)  \right>  \simeq
\frac{(2 \rho_{0}^{2} + \rho_{0})^{1/2}}{(2\pi)^{3/4}}(Dt)^{-1/4}.
\label{rhogeneral}
\end{equation}

It is interesting  that in the low density limit  $\rho_{0} \ll 1$, we
recover the result of Toussaint and Wilczek \cite{22},
\begin{equation}
\left< \rho_{A}(t)  \right> = \left< \rho_{B}(t)  \right> \simeq \frac
{\rho_{0}^{1/2}}{(2 \pi)^{3/4}}(Dt)^{-1/4}\ , \qquad \rho_{0} \ll 1.
\label{rholow}
\end{equation}
However, in the high density limit $\rho_{0} \gg 1$ we obtain the same
decay exponent  $-1/4$ as for  low density but a  different amplitude.
The amplitude now scales as  $\rho_{0}$ in contrast to the low density
limit where it scales like $\rho_{0}^{1/2}$:
\begin{equation}
\left< \rho_{A}(t)  \right> = \left< \rho_{B}(t)  \right> \simeq \frac
{\rho_{0}}{(2 \pi^{3})^{1/4}}(Dt)^{-1/4}\ ,\qquad \rho_{0}\gg 1.
\label{rhohigh}
\end{equation}
Given the  correctness of Eq.(\ref{rhogeneral})  in both low  and high
density limits, we  expect  it to be a good approximation across   the 
whole range of densities.

For   convenience,  let   us  write   Eq.(\ref{rhohigh})   as  $\left<
\rho_{A}(t)   \right>   =   \left<   \rho_{B}(t)  \right>   \simeq   K
\rho_{0}(Dt)^{-\gamma}$.   The analytical values  of $K$  and $\gamma$
are then $K  = 0.35635\ldots$ and $\gamma =  1/4$.  Numerical evidence
has already  substantiated Eq.(\ref{rholow}) \cite{22,24}.   Below, we
present numerical support for our high density calculation.

\begin{figure}
\narrowtext \centerline{\epsfxsize\columnwidth\epsfbox{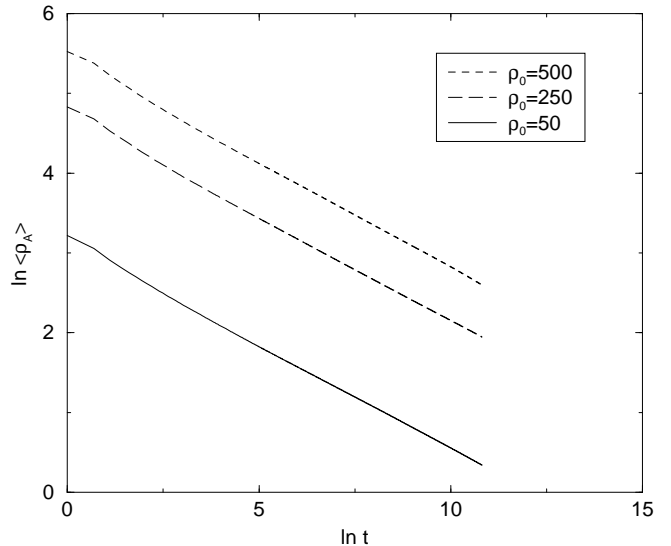}}
\caption{Log-log plot  of the particle  density as a function  of time
for $\rho_{0} = 50, 250 \textrm{ and } 500$.}
\label{Fig1}
\end{figure}

\begin{center}
\begin{tabular}{|c|c|c|} \hline \hline
\hspace{0.3cm} $\rho_{0}$ \hspace{0.3cm} & $\gamma_{num}$
\hspace{0.3cm} & $K_{num}$ \hspace{0.3cm} \\ \hline
\hspace{0.3cm} 50 \hspace{0.3cm} &
\hspace{0.3cm} $0.2576 \pm 0.0005$ \hspace{0.3cm} &
\hspace{0.3cm} $0.3599 \pm 0.0007$ \hspace{0.3cm} \\
\hspace{0.3cm} 250 \hspace{0.3cm} & \hspace{0.3cm} $0.2524 \pm 0.0002$
\hspace{0.3cm} & \hspace{0.3cm} $0.3568 \pm 0.0003$ \hspace{0.3cm} \\
\hspace{0.3cm} 500 \hspace{0.3cm} &
\hspace{0.3cm} $0.2548 \pm 0.0005$ \hspace{0.3cm} &
\hspace{0.3cm} $0.3669 \pm 0.0008$ \hspace{0.3cm} \\ \hline\hline
\end{tabular}
\end{center}

\begin{small}
TABLE 1. Numerical values of the decay exponent $\gamma$ and
the  constant   $K$  where   $\left<  \rho_{A}(t)  \right>   =  \left<
\rho_{B}(t) \right>  \sim K  \rho_{0}(Dt)^{-\gamma}$ for $  \rho_{0} =
50,  250 \textrm  { and  }  500$.  The  numerical values  of $K$  were
obtained by  plotting $\left<\rho_{A}(t) \right> \textrm{  vs. }  t^{-
\gamma}$ and evaluating $K_{num}$ from the gradient of the curve.  The
analytical values are $\gamma = 1/4$ and $K = 0.35645$.
\end{small}

\smallskip

In conducting numerical simulations  of this nature one must naturally
consider  how   to  increment   time.   A  physically   realistic  and
computationally  simple  method is  to  increment  $dt$ by  (1/current
number  of  particles).  However,  the  expediency  of this  technique
clearly does not extend to systems with high densities of particles. A
more   efficient  method   is   to  allow   all   particles  to   move
simultaneously.   In   our  simulations,  therefore,   one  time  step
constitutes  a  jump by  all  particles in  the  system  to a  nearest
neighbour site with  equal probability.  In the low  density limit, we
have  made  comparisons of  the  two  methods  of updating  and  find,
asymptotically that the results are  identical. Within the context of a
low  initial  density,  systems  which permit  multiple  occupancy  of
lattice  sites  generate indistinguishable  results  from those  which
permit a  maximum of one particle  per site. For the  purposes of this
paper,  therefore, all  of the  simulations  are based  on a  parallel
updating  of the particles  and permit  multiple occupancy  of lattice
sites.   Our numerical  simulations  for an  initial  high density  of
particles are performed on a one-dimensional lattice of size $10^{4}$.
Each run is performed for $5  \times 10^{4}$ time steps and we average
our results over 100 runs.   The results are shown for three different
initial densities  in Figure 1. After initial  transients, the log-log
plots  apparently approach straight  lines in  each case.   The values
extracted for  the decay exponent  $\gamma$ and the amplitude  $K$ are
given in Table  1, where the quoted errors  are purely statistical. We
attribute the  small differences between the  measured and theoretical
values to a failure to reach the true asymptotic regime.

\section{TYPE I PERSISTENCE}

Consider the $A +  B \rightarrow \emptyset$ reaction-diffusion process
on  a   one-dimensional  continuum.    The  rate  equations   for  the
concentrations are $\partial_{t} N_{A}  = \partial_{xx} N_{A} - R$ and
$\partial_{t}  N_{B} =  \partial_{xx}  N_{B}  - R$  where  $R$ is  the
reaction rate per unit  volume.  The concentration difference, $\Delta
N =  N_{A}-N_{B}$, obeys  the simple diffusion  equation $\partial_{t}
\phi = \partial_{xx} \phi$. It  is well known that, below the critical
dimension $d_{c}= 4$, the $A  + B \rightarrow \emptyset$ model evolves
to a coarsened  state in which the two  species segregate into domains
of  either $A$  or $B$  particles.  The  domain walls  are  defined by
$\Delta  N  =  0$  and  their  motion is  clearly  determined  by  the
annihilation process  $A +  B \rightarrow \emptyset$.   The persistent
sites in  this model are therefore  defined as those  sites which have
not been touched  by the zeros, $\Delta N=0$,  of the diffusion field,
i.e.\ those  sites which have  never seen an annihilation  process. In
order to  overcome the  discrete nature of  our problem and  model the
diffusive process of the domain walls as described above, we allow our
model to  approach the  continuum limit by  starting with a  very high
initial density of particles, $\rho_{0}  \gg 1$. One might think that,
in  this   limit,  the   entire  system  will   asymptotically  become
nonpersistent. However, we  find this not to be  true. The persistence
decays  according  to $P(t)  \sim  A +  t^{-\theta}$  where  $A$ is  a
constant i.e.\ there are some sites which are always isolated from the
diffusing  boundaries between  domains. We  expect the  offset  $A$ to
vanish  as  $\rho_0 \to  \infty$:  clearly,  in  the real,  continuous
diffusion  problem  $P(t)  \sim  t^{-\theta}$, with  no  offset.   Our
hypothesis  is  that, in  the  limit of  a  large  initial density  of
particles  $  \rho_{0}  \gg 1$,  the  offset  tends  to zero  and  the
persistence exponent is  identical to that which emerges  from a study
of the  diffusion equation in  one dimension, $ \theta  \simeq 0.1207$
\cite{non}.  This  should hold  if the density  during the  entire run
remains large enough $(\rho(t)  \gg 1)$ that the continuum description
in terms  of the diffusion equation  remains valid.  If,  on the other
hand, the simulation  enters the low density regime  at late times, we
expect the approach  of the persistence to its  asymptotic value to be
described by the low-density theory.

In the  limit of  low initial density,  $\rho_{0} \ll 1$,  a realistic
representation of the diffusing  boundaries according to the diffusion
equation  is  not  feasible,  due   to  the  discrete  nature  of  the
problem. In this case it is clear that the entire system cannot become
persistent since, by definition, the  total number of reactions $A + B
\rightarrow \emptyset  \ll L$  where $L$ is  the size of  the lattice.
Hence the relationship  $P(t) \sim A + t^{-\theta}$  is expected, with
$A  \to  1$ as  $\rho_0  \to 0$.  The  large  mean separation  between
particles  implies that  any  site which  experiences an  annihilation
process  is  unlikely to  witness  such  an  event again  i.e.\  every
reaction  process $A  + B  \rightarrow \emptyset$  leads to  a single,
isolated nonpersistent site.  Given  that the total number of reaction
processes is governed by the  initial density of walkers, in the limit
of  sufficiently  low  density  one  can  set a  lower  bound  on  the
persistence $\left<  P( \infty )  \right> = A  \ge 1 -  \rho_{0}$. The
persistence properties  are therefore determined  by the decay  of the
walker density.  The walkers decay according to $\sim t^{-1/4}$ and we
therefore expect that, for $\rho_{0} \ll 1$, $\theta = 1/4$.

Our numerical results are presented  below. The algebraic form for the
persistence  prevents a simple  evaluation of  the exponent.   A three
parameter  nonlinear curve  of the  form $A  +  Bt^{-\theta_{nl}}$ was
fitted  to the  data to  ascertain the  value $\theta_{nl}$  where the
subscript  denotes the  implementation  of a  nonlinear curve  fitting
technique.  The fit  was  carried  out using  $t$  as the  independent
variable, though the  data are plotted against $\ln  t$ for clarity of
presentation.  To demonstrate the  validity of determining $\theta$ in
this manner, we  also plot our data in the  form $\left< P(t) \right>$
against $t^{-  \theta_{nl}}$, where the correct value  of $\theta$, in
such a  plot, manifests itself as  a straight line  graph. An unbiased
method of  determining the exponent is also  presented, by numerically
differentiating the data and displaying the results on a log-log plot.
The value  of $\theta$ determined  in this manner is  denoted $\theta_
{diff}$.  We  consider the  high density limit  first and take  as our
initial values $\rho_{0}  = 500 \textrm{ and }  250$.  The simulations
are performed  on a one-dimensional  lattice of size $10^{4}$  and run
for  $5 \times 10^{4}$  time steps.  We average  our results  over 100
runs.  In  the nonlinear  fits, the  regime $t \le  3 \times  10^4$ is
fitted, corresponding to $\ln t \le 10.3$.

\begin{figure}
\narrowtext \centerline{\epsfxsize\columnwidth\epsfbox{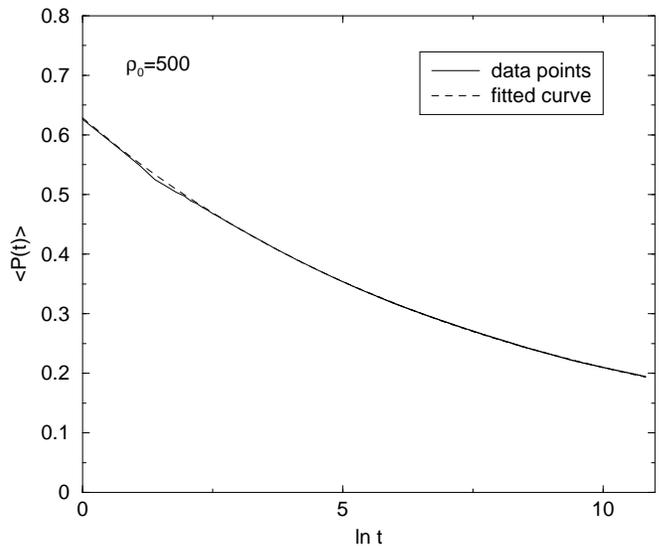}}
\caption{A nonlinear fit of the form $A + Bt^{-\theta_{nl}}$ is made 
to the fraction of persistent sites at time $t$ for an initial density
$\rho_{0}  =  500$. The values of the fit parameters are $A  =  0.05  
\pm 0.02$,  $B  =  0.57 \pm  0.005$, and $\theta_{nl} = 0.13 \pm 0.01$.}
\label{Fig2}
\end{figure}

The data  are presented in Figures  2-7. Figure 2  shows the nonlinear
fit for  $\rho_0=500$, from  which the exponent  $\theta_{nl}=0.13 \pm
0.01$  is  extracted, consistent  with  the  diffusion value,  $\theta
\simeq  0.1207$.  The  plot,  in Figure  3,  of $\langle  P(t)\rangle$
against  $t^{-0.129}$   (where  $0.129$  is  the   best-fit  value  of
$\theta_{nl}$ from Figure 2) yields a good straight line at large $t$.
The   unbiased   determination    of   $\theta$   from   a   numerical
differentiation  of  the  data  is   shown  in  Figure  4,  where  the
logarithmic derivative  of $\langle  P(t) \rangle$ is  plotted against
$t$ on a  log-log plot.  The resulting data should  be a straight line
of   slope  $-\theta$.    The   best-fit  value   of   the  slope   is
$-\theta_{diff}=-0.12  \pm 0.01$.  The  slope was  extracted from  the
region between  the arrows, where initial transients  have decayed but
the  data are not  yet too  noisy. The  expected slope,  $-0.1207$, is
shown as a guide to the eye.

\begin{figure}
\narrowtext \centerline{\epsfxsize\columnwidth\epsfbox{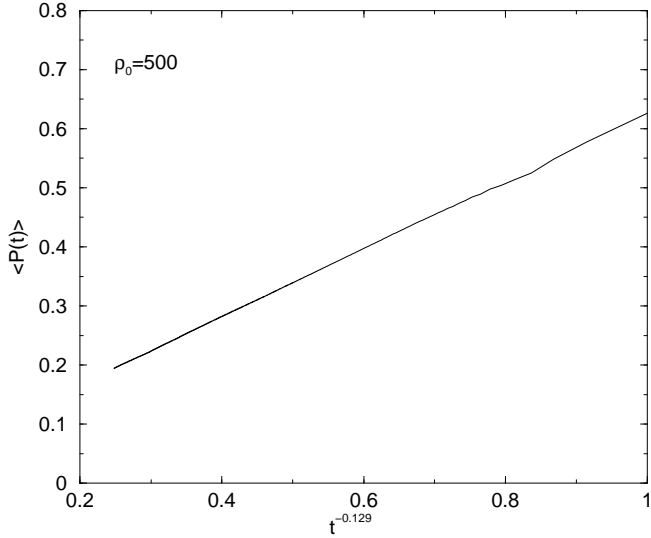}}
\caption{The fraction  of  persistent   sites   is  plotted   against
$t^{-\theta_{nl}}$ for  $\rho_{0} = 500$. The value  $\theta_{nl} =
0.129$ is taken from the nonlinear curve fit in Fig.2.}
\label{Fig3}
\end{figure}

\begin{figure}
\narrowtext \centerline{\epsfxsize\columnwidth\epsfbox{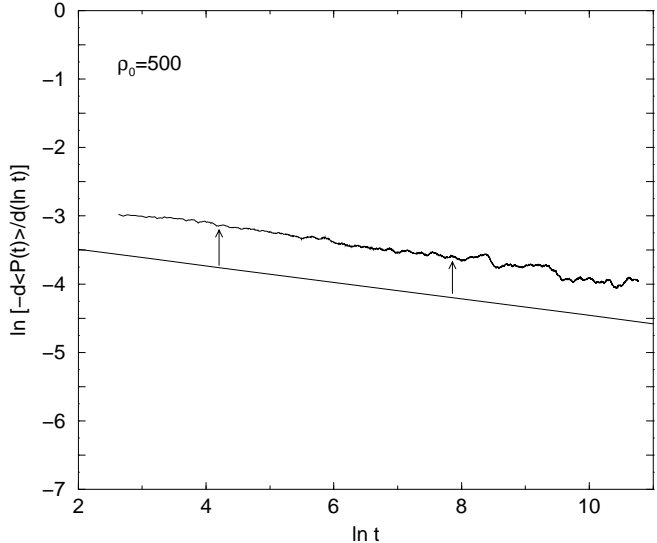}}
\caption{The fraction  of persistent sites at time  $t$ is numerically
differentiated ($\Delta(\ln t) = 0.1$) and presented on a log-log plot
for $\rho_{0}  = 500$.  The gradient, taken  between the arrows, gives  
$\theta_{diff}= 0.12 \pm  0.01$. The straight  line has slope -0.1207.}
\label{Fig4}
\end{figure}

The equivalent data for $\rho_0=250$ are shown in Figures 5-7. In this
case one obtains $\theta_{nl} =  \theta_{diff} = 0.13 \pm 0.01$, again
consistent with the diffusion equation result.
 
\begin{figure}
\narrowtext \centerline{\epsfxsize\columnwidth\epsfbox{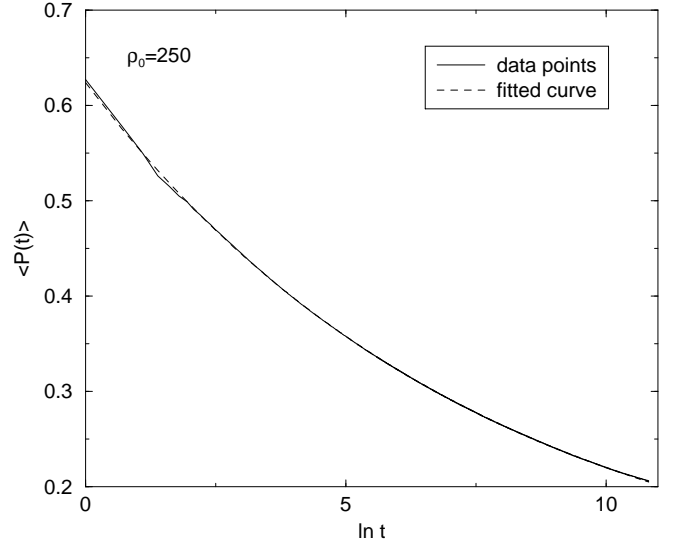}}
\caption{Same as  Fig.2, but for $\rho_{0}  = 250$. The  values of the
fit  parameters  are $A  =  0.07  \pm 0.02$,  $B  =  0.55 \pm  0.005$,
$\theta_{nl} = 0.13 \pm 0.01.$}
\label{Fig5}
\end{figure}

\begin{figure}
\narrowtext \centerline{\epsfxsize\columnwidth\epsfbox{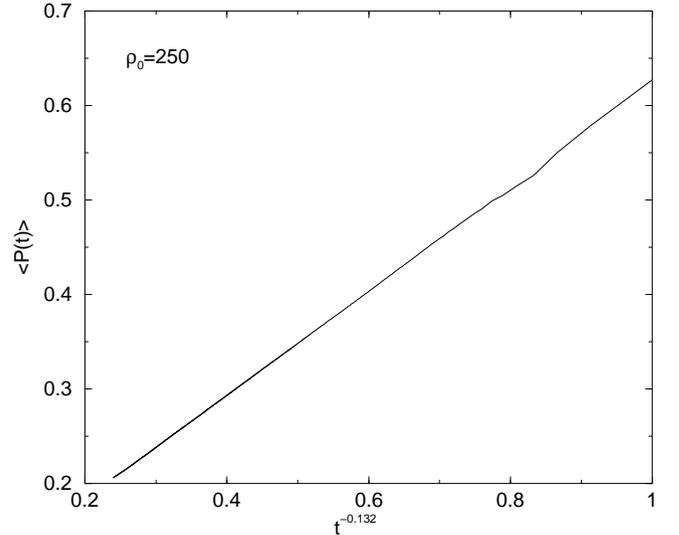}}
\caption{Same as Fig.3, but for  $\rho_{0} = 250$, with $\theta_{nl} =
0.132.$}
\label{Fig6}
\end{figure}

Notice that  the values of $A$  obtained from the  non-linear fits are
quite small -- $0.05(2)$ and  $0.07(2)$ for $\rho_0=500$ and $\rho_0 =
250$ respectively. An equivalent  fit for $\rho_0=50$ gives the larger
value $A=0.17(2)$, so the data are consistent with the hypothesis that
$A \to  0$ for  $\rho_0 \to  \infty$, although the  data are  not good
enough to extract the functional dependence of $A$ on $\rho_0$ in this
limit.

\begin{figure}
\narrowtext \centerline{\epsfxsize\columnwidth\epsfbox{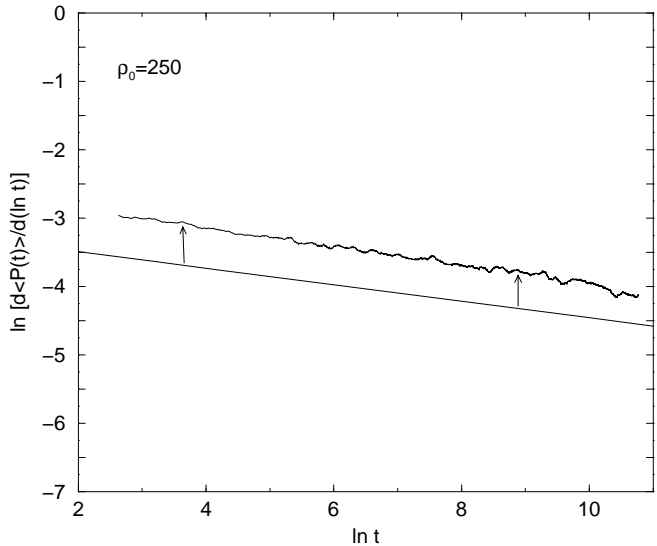}}
\caption{Same as Fig.4, but for $\rho_{0} = 250$. $\theta_{diff} =
0.13 \pm 0.01$.}
\label{Fig7}
\end{figure}

\begin{figure}
\narrowtext \centerline{\epsfxsize\columnwidth\epsfbox{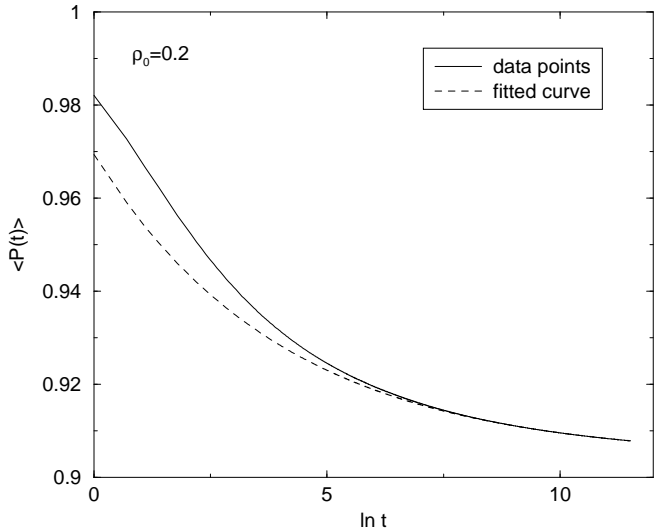}}
\caption{Same as Fig.2, but for  $\rho_{0} = 0.2$.  The fit parameters
are $A = 0.90 \pm 0.02$, $B = 0.06 \pm 0.005$, $\theta_{nl} = 0.25 \pm
0.01$.}
\label{Fig8}
\end{figure}

Our low density results are  presented below. We choose as our initial
densities $\rho_{0}  = 0.2  \textrm{ and }  0.1$. In this  regime, our
simulations are  performed on a lattice  of size $10^{5}$  and run for
$10^{5}$ time steps.  The system  takes longer to enter the asymptotic
regime due  to the low  initial density and  this is reflected  in our
extended number  of time steps. We  average our results  over 50 runs.
Figures 8 and  11 show the same type of  non-linear fit, $\langle P(t)
\rangle = A + B/t^{\theta_{nl}}$ that was used in Figures 2 and 5. The
fit works well except at early  times, and the fitted values are close
to $\theta=1/4$  as anticipated. Note  that the logarithmic  scale for
the abscissa greatly expands  the early-time regime. The nonlinear fit
was restricted to the range $10^4 <  t < 10^5$ (i.e.\ the last 90\% of
each run),  corresponding to $9.2 < \ln  t < 11.5$.  Figures  9 and 12
show  the  same  data   plotted  as  $\langle  P(t)  \rangle$  against
$t^{-\theta_{nl}}$, and  reveal the expected linear  behaviour at late
times.   Finally, Figures 10  and 13  give the  log-log plots  for the
differentiated data, from which the exponent estimates $\theta_{diff}$
are obtained.  The  slope is measured between the  arrows in Figs.\ 10
and  13,  corresponding to  a  similar region  to  that  used for  the
non-linear fits.

\begin{figure}
\narrowtext \centerline{\epsfxsize\columnwidth\epsfbox{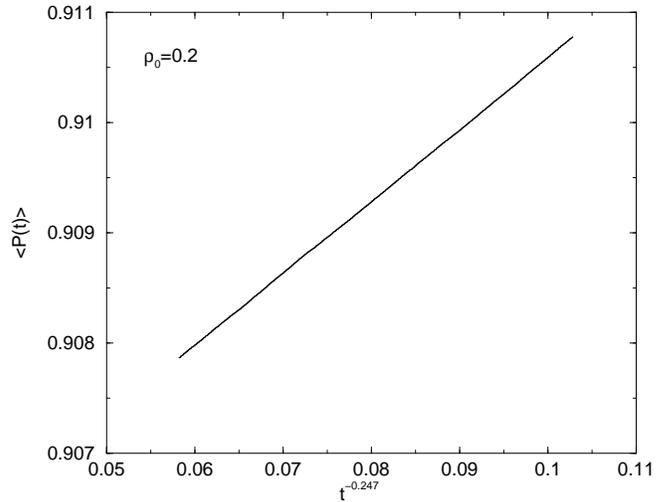}}
\caption{Same  as Fig.3,  but  for $\rho_{0}  =  0.2$. $\theta_{nl}  =
0.247$.}
\label{Fig9}
\end{figure}

\begin{figure}
\narrowtext \centerline{\epsfxsize\columnwidth\epsfbox{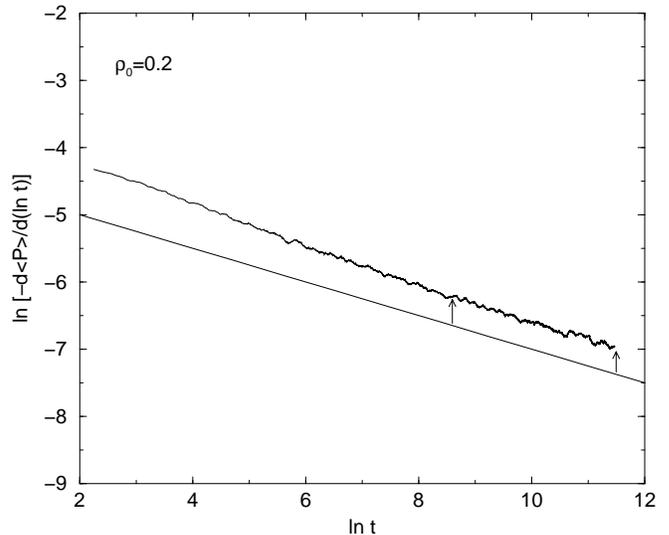}}
\caption{Same as  Fig.4, but for  $\rho_{0} = 0.2$. Here  the straight
line  has a  gradient of  -1/4.  The  gradient of  the  data, measured
between the arrows, is $\theta_{diff} = 0.25 \pm 0.01$.}
\label{Fig10}
\end{figure}

The  estimates $\theta_{nl}$  and $\theta_{diff}$  of  the persistence
exponent, for  both low  and high density  regimes, are  summarised in
Table  2. A natural  consequence of  the slow  algebraic decay  of the
particle density is that the  asymptotic regime, in the case $\rho_{0}
\ll  1$, occurs  only  after many  time  steps. We  expect the  result
$\theta \simeq 1/4$ for the low-density regime to manifest itself more
clearly in systems run for a greater number of time steps.

\begin{figure}
\narrowtext \centerline{\epsfxsize\columnwidth\epsfbox{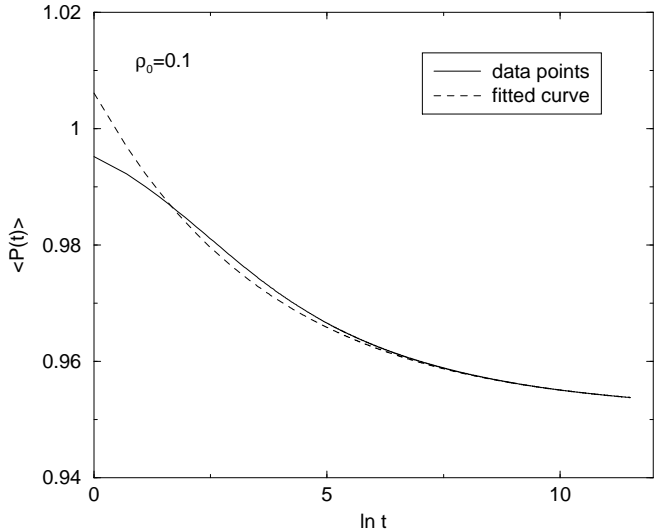}}
\caption{Same as Fig.2, but for  $\rho_{0} = 0.1$.  The fit parameters
are $A = 0.95 \pm 0.02$, $B = 0.05 \pm 0.005$,  $\theta_{nl} = 0.26 \pm
0.01$.}
\label{Fig11}
\end{figure}

\begin{figure}
\narrowtext \centerline{\epsfxsize\columnwidth\epsfbox{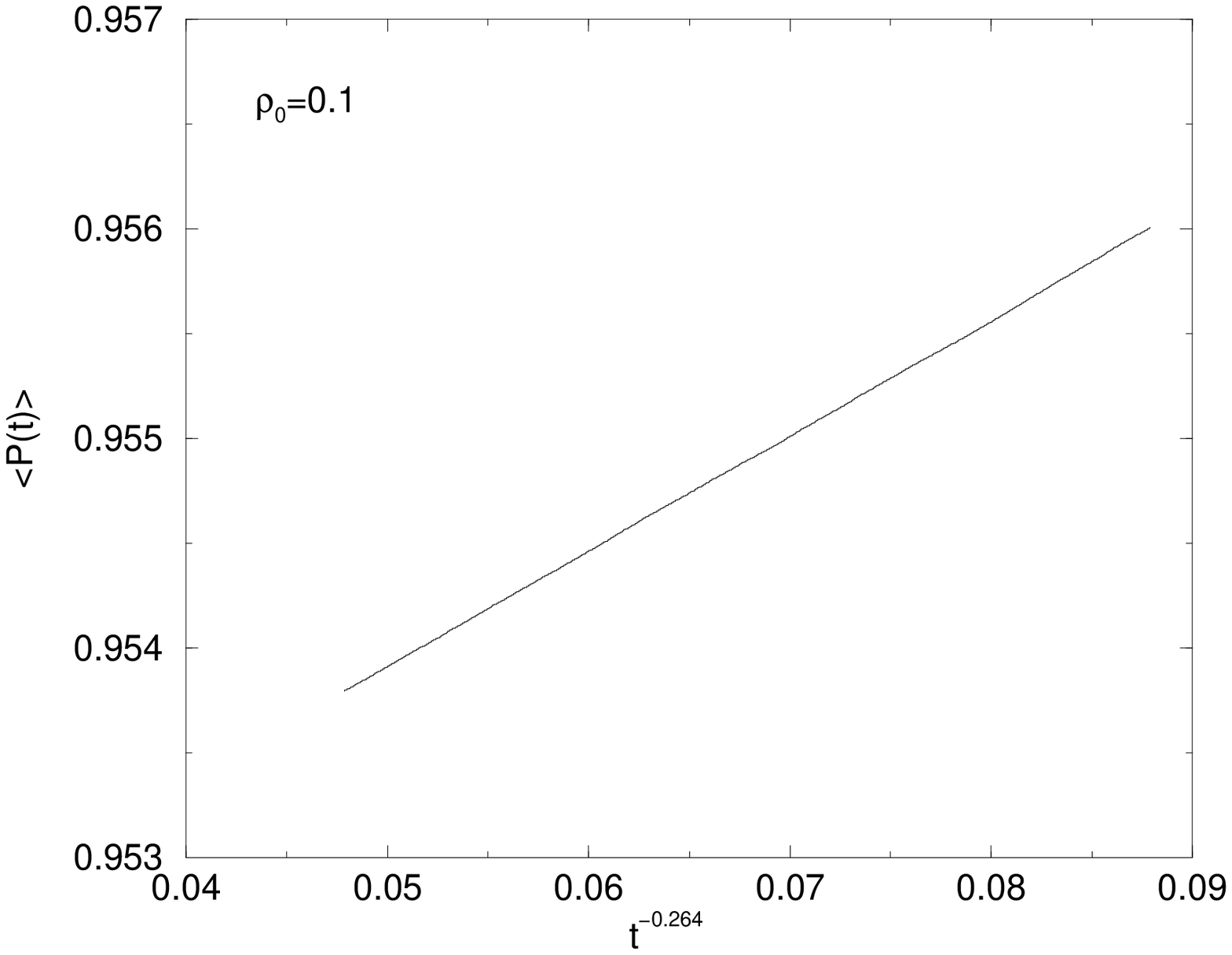}}
\caption{Same  as  Fig.3, but for  $\rho_{0}  =  0.1$.  $\theta_{nl}  =
0.264$.}
\label{Fig12}
\end{figure}

\bigskip

\begin{tabular}{|c|c|c|c|c|}
\hline\hline $\rho_{0}$ & $A$  & $B$ & $\theta_{nl}$ & $\theta_{diff}$
\\ \hline 500 & $0.05 \pm 0.02$ & $0.57 \pm 0.005$ & $0.13 \pm 0.01$ &
$0.12 \pm 0.01$  \\ 250 & $0.07  \pm 0.02$ & $0.55 \pm  0.005$ & $0.13
\pm 0.01$  & $0.13  \pm 0.01$  \\ 0.2 &  $0.90 \pm  0.02$ &  $0.06 \pm
0.005$ & $0.25 \pm 0.01$ & $0.25  \pm 0.01$ \\ 0.1 & $0.95 \pm 0.02$ &
$0.05 \pm 0.005$ & $0.26 \pm 0.01$ & $0.27 \pm 0.01$ \\ \hline
\end{tabular}

\smallskip

\begin{small}
TABLE  2.   Values  of  the  exponent  $\theta$  in  the  relationship
$P(t)=A+Bt^{-\theta}$, where $\theta$  has been evaluated according to
a  nonlinear curve  fitting  technique ($\theta_{nl}$)  and through  a
method  of numerical  differentiation  ($\theta_{diff}$).  We  expect,
asymptotically, $\theta \simeq 0.1207$ for $\rho_{0}\gg 1$ and $\theta
= 1/4$ for $\rho_{0} \ll 1$.
\end{small}

\begin{figure}
\narrowtext \centerline{\epsfxsize\columnwidth\epsfbox{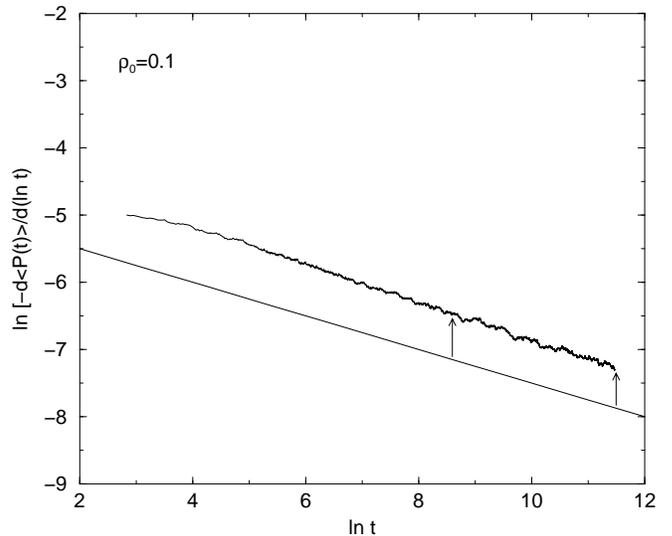}}
\caption{Same as  Fig.10 for $\rho_{0}  = 0.1$. $\theta_{diff}  = 0.27
\pm 0.01$.}
\label{Fig13}
\end{figure}

\section{TYPE II PERSISTENCE}

In  this  section we  address  the  following  question: What  is  the
asymptotic probability $P(t)$ that a given site has never been visited
by an $A$ or $B$ particle, in  a one dimensional system of $A$ and $B$
random  walkers   subject  to  the  $A  +   B  \rightarrow  \emptyset$
reaction-diffusion process? Necessarily,  we pose this question within
the context of a low initial density.

Our  approach  involves  the   introduction  of  a  toy  model,  whose
predictions we will compare  to simulation results. The model consists
of  a  system  of  noninteracting diffusing  particles,  with  initial
density  $\rho_{0}  =  N(0)/L$,  on  a  one-dimensional  lattice  with
diffusion constant $D=1/2$. Note  that the definition of $\rho_{0}$ in
type II persistence refers to  the total density of initial particles,
in contrast to  type I persistence, where $\rho_{0}$  referred to half
the total density.

Particle decay is invoked by allowing each particle to vanish at every
time step  with some (usually time-dependent)  probability.  We define
$f(t)$ to  be the  fraction of particles  remaining at time  $t$ i.e.\
$f(t) = \rho(t)/ \rho_{0}$. We are then free to mimic the behaviour of
any reaction-diffusion process by a suitable choice of $f(t)$.

The analysis of persistence in  this toy model is straightforward.  We
first define the  probability, $Q(x,t)$, that the origin  has not been
crossed by a given diffusing particle with initial position $x>0$.  We
define our lattice to be the interval $(-L/2,L/2)$. Then an elementary
calculation  gives $Q(x,t)  = \textrm{erf}(x/2  \sqrt{Dt})$,  where we
have assumed that $x \ll L$ and $L^2 \gg Dt$, i.e. we have effectively
taken  the limit $L  \to \infty$  in the  calculation of  $Q(x,t)$.  A
simple approach in  the calculation of the number  of persistent sites
is to  consider the first-passage-time  distribution $P_{1}(t)$, where
$P_{1}(t)dt$ is the probability that  the first crossing of the origin
occurs  in $(t,t+dt)$. Then  $P_{1}(t)dt =  Q(x,t) -  Q(x,t+dt)$ gives
$P_1(t) = -dQ/dt$.  Since the  diffusing particle survives to time $t$
with  probability  $f(t)$,  the  probability  for  the  origin  to  be
persistent at time $t$ is given by,
\begin{eqnarray}
Q(x,t) & =  & 1 - \int_{0}^{t}dt'P_{1}(t')f(t') \nonumber\\ &  = & 1 -
\frac{x}{2  \sqrt{D  \pi}}  \int_{0}^{t}  \frac{ds}{s^{3/2}}f(s)  \exp
\left( - \frac{x^{2}}{4Ds} \right).
\end{eqnarray}

In the presence of many random walkers, whose random initial positions
have a uniform distribution over  space, the mean persistence is given
by $\left< P(t) \right> = \left< Q(|x|,t) \right> ^{\rho_{0} L}$ where
$\rho_{0} L =  N$ the total number of initial  walkers and the average
is  over  the  initial  position  of  a  walker,  assumed  uniform  in
$(-L/2,L/2)$. This gives
\begin{eqnarray}
\left< P(t) \right> & = &  \left[ 1- \frac{1}{2L \sqrt{ D \pi}} \int_{
- \frac{L}{2}}^{  \frac{L}{2}}  dx|x| \int_{0}^{t}  \frac{ds}{s^{3/2}}
f(s)   \right.   \nonumber\\   &&   \left.   \times   \exp  \left(   -
\frac{x^{2}}{4Ds} \right) \right] ^{\rho_{0} L}.
\end{eqnarray}
For $L \rightarrow \infty$ we obtain,
\begin{equation}
\left< P(t) \right> = \exp  \left( -2 \rho_{0} \sqrt { \frac{D}{ \pi}}
\int_{0}^{t} \frac{ds}{s^{1/2}}f(s) \right).
\label{pii}
\end{equation}

\begin{figure}
\narrowtext \centerline{\epsfxsize\columnwidth\epsfbox{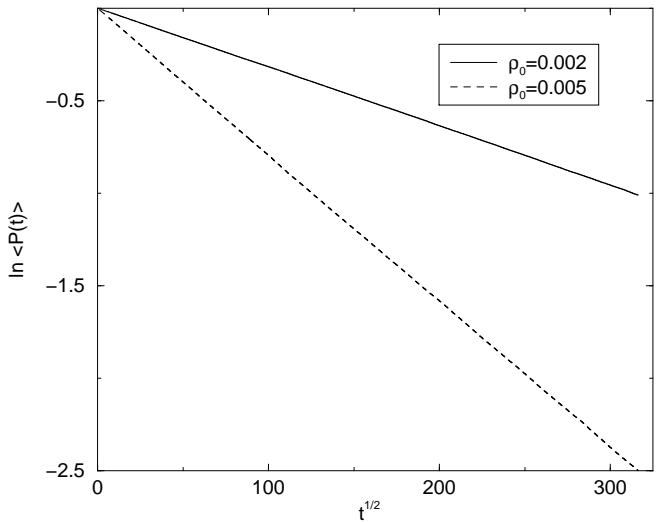}}
\caption{Log-linear  plot   of  the   persistence  for  a   system  of
noninteracting diffusing particles with no decay. }
\label{Fig14}
\end{figure}

A system of strictly diffusing  particles with no decay is modelled by
setting  $f(s)  =  1$. In  this  case,  $\left<  P(t) \right>  =  \exp
(-At^{1/2})$  where   $A  =  4  \rho_{0}  \sqrt{D/   \pi}$.   In  Fig.
\ref{Fig14}   we    present   a   numerical    verification   of   our
calculation. The gradient of the  graphs gives $-A$. The values of $A$
extracted from the data are compared with the numerical predictions in
Table 3. The agreement is excellent.

\bigskip

\begin{center}
\begin{tabular}{|c|c|c|}
\hline\hline $\hspace{0.5cm}\rho_{0}\hspace{0.5cm} $ &
\hspace{0.5cm} $A_{th}$ \hspace{0.5cm} &
\hspace{0.5cm} $A_{num}$ \hspace{0.5cm} \\ \hline
\hspace{0.5cm} 0.002 \hspace{0.5cm} &
\hspace{0.5cm} 0.00319 \hspace{0.5cm} &
\hspace{0.5cm} $0.00320(5)$ \hspace{0.5cm} \\
\hspace{0.5cm} 0.005 \hspace{0.5cm} &
\hspace{0.5cm} 0.00797 \hspace{0.5cm} &
\hspace{0.5cm} $0.00790(5)$ \hspace{0.5cm} \\ \hline
\end{tabular}
\end{center}

\begin{small}
TABLE  3.  The persistence  properties of  a system  of noninteracting
particles with zero probability of  decay is described by $\left< P(t)
\right> \sim \exp (-At^{1/2})$.  Above  we provide a comparison of the
values of $A$  as deduced from the theory  $A_{th}$ and from numerical
simulations $A_{num}$.
\end{small}

\smallskip

Our model can also be applied  to the $q$-state Potts model, which has
$q$  distinct but  equivalent  ordered phases.   The $A+A  \rightarrow
\emptyset$  model  corresponds to  $q=2$,  while  $A+A \rightarrow  A$
corresponds to  $q= \infty$.  The density  of walkers in  the 1D Potts
model is known to decay asymptotically according to \cite{27},
\begin{equation}
\rho(t) \simeq \frac{q-1}{q} \frac{1}{\sqrt{2 \pi Dt}}.
\label{rho(q)}
\end{equation}
We can derive a generalised value  of $\theta(q)$ for our toy model by
substituting  $f(s) =  \rho(s) /  \rho_{0}$ into  Eq.(\ref{pii}) where
$\rho(s)$ is given by Eq.(\ref{rho(q)}). This gives,
\begin{equation}
\left< P(t) \right> \sim t^{- \theta_T(q)}\ ,
\label{powerlaw}
\end{equation}
where
\begin{equation}
\theta_T(q) = \frac{ \sqrt{2}}{ \pi} \left( \frac{q-1}{q} \right)\ ,
\label{thetaii}
\end{equation}
and the subscript $T$ denotes `toy'.  Despite the simplistic nature of
our  model, a power  law decay  for the  persistence is  predicted, in
agreement  with  known   results.   Our  expression  for  $\theta(q)$,
(Eq.(\ref{thetaii}))  is clearly  a  poor approximation  to the  exact
expression for the  1D Potts model obtained by  Derrida \textit{et al}
\cite{12}.
\begin{equation}
\theta_P(q)       =       -\frac{1}{8}       +       \frac{2}{\pi^2}\,
\left[\cos^{-1}\left(\frac{2-q}{\sqrt{2}\,q}\right)\right]^2\ ,
\label{theta(q)} 
\end{equation}
where  the   subscript  $P$  denotes  `Potts'.   The   values  of  the
persistence  exponent  returned  by   our  model  for  the  $q=2$  and
$q=\infty$ cases are,  $\theta_T(q=2) = 0.225$ and $\theta_T(q=\infty)
= 0.450$. The values given  by Eq.(\ref{theta(q)}) for the Potts model
are  $\theta_P(q=2)  =  3/8$  and  $\theta_P(q=  \infty)  =  1$.   The
encouraging feature of  our toy model, however, is  that the essential
dynamics  of  the  system  leading  to  the  power-law  decay  of  the
persistence  in  Eq.(\ref{powerlaw}) is  correctly  identified as  the
$t^{-1/2}$ decay  in the  number of surviving  random walkers  at time
$t$.  If the  number of surviving walkers decays  as $t^{-\alpha}$ for
large  $t$,  the toy  model  predicts,  via  Eq.(\ref{pii}), that  the
persistence  will  decay as  a  stretched  exponential, $\langle  P(t)
\rangle \sim(-At^{1/2 - \alpha})$,  for $\alpha < 1/2$, while $\langle
P(t)  \rangle$  will  approach  a  non-zero  constant  for  $\alpha  >
1/2$. The borderline case, $\alpha=1/2$, yields power-law decay of the
persistence, as we have seen.

We  naturally  now  turn our  attention  to  the  $A +  B  \rightarrow
\emptyset$ reaction process  to see if the correct  time dependence of
the persistence also  emerges from our model.  The  density of $A$ and
$B$ particles in this case  in given by Eq.  (\ref{rholow}). Therefore
$\left< \rho(t) \right>$ is given by,
\begin{equation}
\left<     \rho(t)    \right>    \sim     \frac    {\rho_{0}^{1/2}}{(2
\pi^{3})^{1/4}}(Dt)^{-1/4}\ , \qquad \rho_{0} \ll 1\ ,
\label{rho(t)}
\end{equation}
where $\rho(t)  = \rho_{A}(t) + \rho_{B}(t)$ and  $\rho_{0} = N(0)/L$.
Substituting $f(s)  = \rho(s)/ \rho_{0}$  where $\rho(s)$ is  given by
Eq.(\ref{rho(t)}) into Eq.(\ref{pii}) yields $\left< P(t) \right> \sim
\exp (-Bt^{1/4})$ with  $B = (2^{11}D/ \pi^{5})^{1/4} \rho_{0}^{1/2}$.
For the  $A+B \rightarrow \emptyset$  reaction we therefore  predict a
stretched exponential decay of the persistence, with exponent $1/4$.

\begin{figure}
\narrowtext \centerline{\epsfxsize\columnwidth\epsfbox{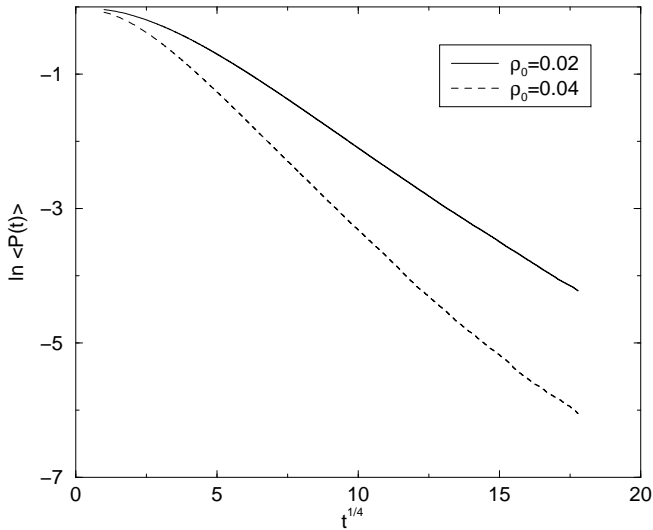}}
\caption{Log-linear plot of the persistence for the $A + B \rightarrow
\emptyset$ reaction-diffusion process.  $\left< P(t) \right> \sim \exp
(-Bt^{\gamma})$ with $\gamma=1/4$. }
\label{Fig15}
\end{figure}

The  results of  our numerical  simulations are  presented  below. The
simulations  are  performed  on  a  one-dimensional  lattice  of  size
$10^{5}$ for $10^{5}$  time steps. We choose as  our initial densities
$\rho_{0}  =  0.02$ and  $\rho_{0}  = 0.04$.   A  direct  test of  the
stretched exponential  prediction is obtained  by plotting $\ln\langle
P(t) \rangle$  against $t^{1/4}$, as in  Figure 15. The  fact that the
data  do not  clearly show  the expected  straight line  behaviour may
indicate that  the density is not yet  in the regime where  it is well
described by the $t^{-1/4}$ form given by Eq.(\ref{rho(t)}). Indeed, a
direct study of the density indicates that the $t^{-1/4}$ behaviour is
only  evident at  the latest  times  reached in  the simulations.   An
alternative analysis involves differentiating the data with respect to
$\ln t$ and presenting the result  as a double logarithmic plot, as in
Figures 16 and  17. Here there is evidence that  the data approach the
expected slope of  $1/4$ at late times, though  the noisy character of
the differentiated data at the  very latest times tends to obscure the
asymptotic behaviour.

\begin{figure}
\narrowtext \centerline{\epsfxsize\columnwidth\epsfbox{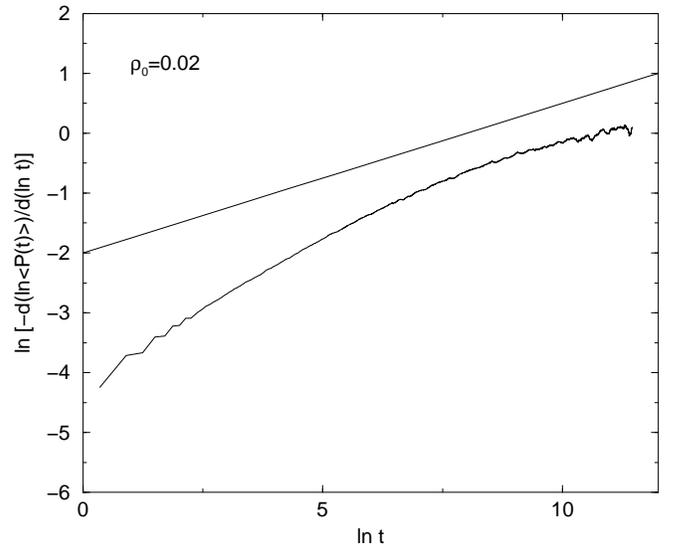}}
\caption{The log of the persistent fraction at time $t$ is numerically
differentiated $(\Delta(\ln t) = 0.1)$ and presented on a log-log plot
for  $\rho_{0}=0.02$ The  gradient of  the curve  gives  $\gamma$. The
straight line has a slope of $1/4$. }
\label{Fig16}
\end{figure}
\begin{figure}
\narrowtext \centerline{\epsfxsize\columnwidth\epsfbox{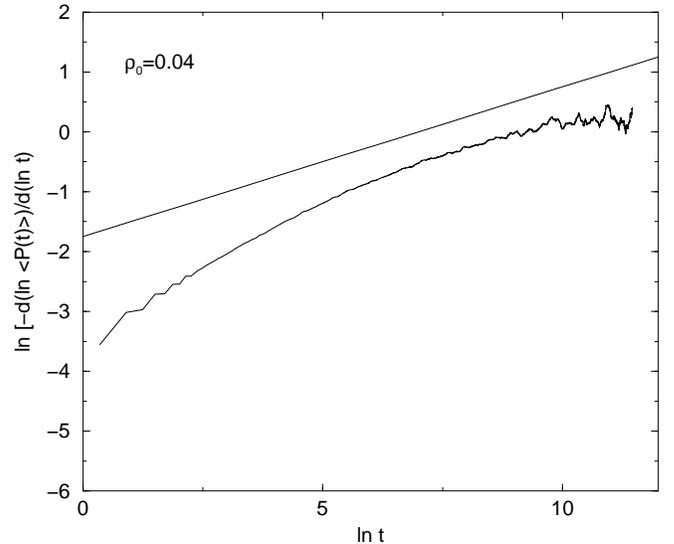}}
\caption{Same as Fig.16 but for $\rho_{0} = 0.04$. }
\label{Fig17}
\end{figure}

In short, the data as presented in Figures 15, 16 and 17 are not clean
enough to  provide a  convincing test of  our prediction  $\left< P(t)
\right>  \sim \exp  (-Bt^{\gamma})$  with $\gamma  =  1/4$.  The  slow
approach  of  the  particle   density  to  its  asympotic  $1/t^{1/4}$
behaviour is, in  our view, the reason for this. A  better test of our
model, therefore, is to consider  the persistence as a function of the
actual running density of  particles, rather than using the asymptotic
behaviour  of  the  density  in  Eq.(\ref{pii}).  Taking  the  log  of
Eq.(\ref{pii}) and differentiating with respect to $\ln t$ yields,
\begin{equation}
-\frac{d\ln  \left< P(t)  \right>}{d \ln  t} =  2 \sqrt{\frac{D}{\pi}}
\rho(t)t^{1/2}\ ,
\label{plot}
\end{equation}
since $\rho_0f(t)=\rho(t)$, the running density. The left-hand side of
Eq.(\ref{plot})    and    $\rho(t)t^{1/2}$    (without   the    factor
$2\sqrt{D/\pi}$)  are  plotted  in  Figures  18 and  19  for  the  two
densities studied.   The two curves  agree rather well over  the whole
range of $\ln t$, except at  the latest times, where the date are very
noisy.   The factor $2\sqrt{D/\pi}$ in Eq.(\ref{plot}),   omitted from  
the plots, has the numerical  value $\sqrt{2/\pi} \simeq  0.8$ (recall  
$D=1/2$),  whereas  Figures 18 and 19  suggest  this  number should be 
closer to unity. Given the crudeness  of the  toy model, however,  the 
agreement  between the data and the model is surprisingly good.

\begin{figure}
\narrowtext
\centerline{\epsfxsize\columnwidth\epsfbox{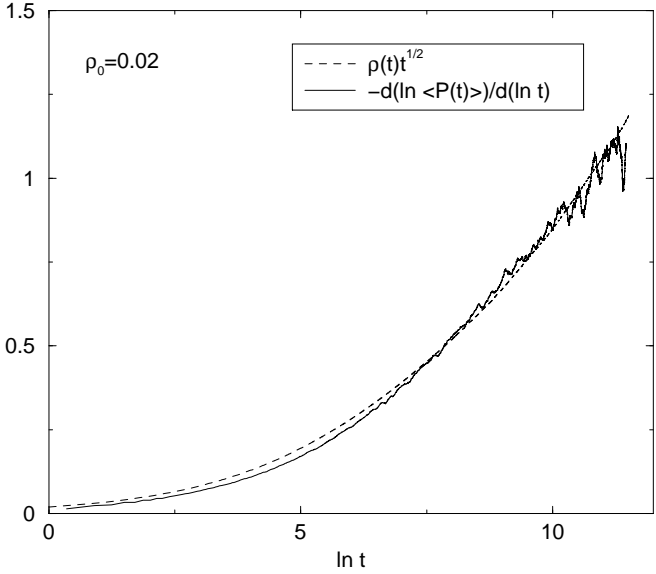}}
\caption{A direct test of Eq.(\ref{plot}) for $\rho_0 =0.02$.}
\label{Fig18}
\end{figure}
\begin{figure}
\narrowtext
\centerline{\epsfxsize\columnwidth\epsfbox{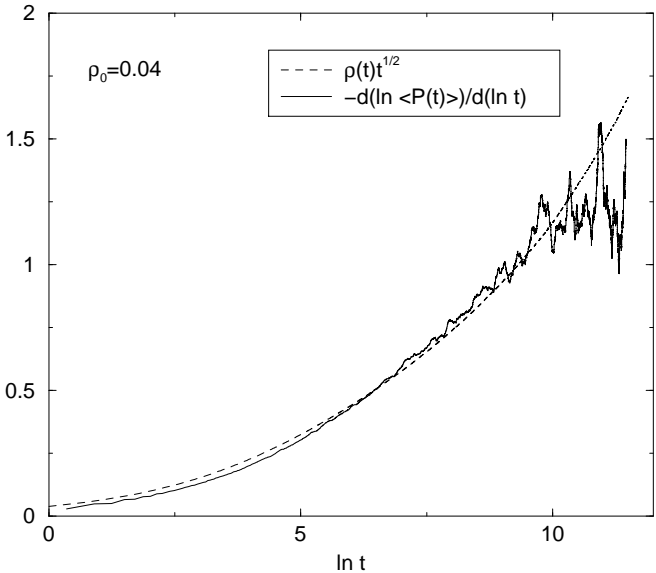}}
\caption{Same as Figure 18, but for $\rho_0 = 0.04$.}
\label{Fig19}
\end{figure}

\section{SUMMARY}

In this paper  we have investigated the persistence  properties of the
$A  + B  \rightarrow \emptyset$  reaction-diffusion process  using two
distinct definitions of persistence. In type I persistence, we studied
the  fraction  of sites  which  had  never  witnessed an  annihilation
process.  In  the high density limit  $\rho_{0} \gg 1$  we argued that
the persistence exponent is that which emerges from a study of the one
dimensional  diffusion equation,  $\partial_{t}  \phi =  \partial_{xx}
\phi$,  where $\theta  \simeq 0.1207$  \cite{non}, and  presented data
consistent with this  result.  In the low density  limit $\rho_{0} \ll
1$ we argued that the  persistence properties are governed by the same
exponent  that describes  the decay  of the  particle  density, giving
$\theta =  1/4$. Again, the data  support this prediction.  In type II
persistence,  we  considered  the  probability that  a  site  remained
unvisited  by either an  $A$ or  $B$ particle.  Our approach,  in this
case,  was  to develop  a  toy  model, which  can  be  applied to  any
reaction-diffusion process and expresses the persistence properties in
terms  of the  density  of  the reactants.  For  the $A+B  \rightarrow
\emptyset$  process  the  model  predicts $\left<  P(t)  \right>  \sim
\exp(-Bt^{1/4})$.  The  data  are  consistent with  this  result  when
allowance  is made  for  the actual  time-dependence  of the  particle
density (rather  than just its  asymptotic form). An obvious  goal for
future study would be to  to place this stretched-exponential decay on
a firmer theoretical foundation.

\section{ACKNOWLEDGMENT}
This work was supported by EPSRC (UK).

\end{multicols}

\end{document}